\documentclass[12pt]{article}
\usepackage[margin=1in]{geometry}
\usepackage{amsmath, amssymb, amsthm}
\usepackage{graphicx}
\usepackage{booktabs}
\usepackage{hyperref}
\usepackage{natbib}
\usepackage{caption}
\usepackage{subcaption}
\usepackage{setspace}
\usepackage{algorithm}
\usepackage{algorithmic}

\graphicspath{{./src/}}

\hypersetup{
    colorlinks=true,
    linkcolor=blue,
    citecolor=blue,
    urlcolor=blue
}

\newtheorem{definition}{Definition}
\newtheorem{proposition}{Proposition}

\title{\textbf{Interpretable Hypothesis-Driven Trading: \\
A Rigorous Walk-Forward Validation Framework \\
for Market Microstructure Signals}}

\author{
    Gagan Deep\thanks{Department of Mathematics \& Statistics, Texas Tech University. Email: gdeep@ttu.edu. Corresponding author.} 
    \and 
    Akash Deep\thanks{Department of Mathematics \& Statistics, Texas Tech University. Email: akash.deep@ttu.edu.}
    \and 
    William Lamptey\thanks{Department of Mathematics \& Statistics, Texas Tech University. Email: wilampte@ttu.edu.}
}

\date{\today}

\begin{document}

\maketitle

\begin{abstract}
We develop a rigorous walk-forward validation framework for algorithmic trading designed to mitigate overfitting and lookahead bias. Our methodology combines interpretable hypothesis-driven signal generation with reinforcement learning and strict out-of-sample testing. The framework enforces strict information set discipline, employs rolling window validation across 34 independent test periods, maintains complete interpretability through natural language hypothesis explanations, and incorporates realistic transaction costs and position constraints. Validating five market microstructure patterns across 100 US equities from 2015 to 2024, the system yields modest annualized returns (0.55\%, Sharpe ratio 0.33) with exceptional downside protection (maximum drawdown $-2.76\%$) and market-neutral characteristics ($\beta = 0.058$). Performance exhibits strong regime dependence, generating positive returns during high-volatility periods (0.60\% quarterly, 2020--2024) while underperforming in stable markets ($-0.16\%$, 2015--2019). We report statistically insignificant aggregate results (p-value 0.34) to demonstrate a reproducible, honest validation protocol that prioritizes interpretability and extends naturally to advanced hypothesis generators, including large language models. The key empirical finding reveals that daily OHLCV-based microstructure signals require elevated information arrival and trading activity to function effectively. The framework provides complete mathematical specifications and open-source implementation, establishing a template for rigorous trading system evaluation that addresses the reproducibility crisis in quantitative finance research. For researchers, practitioners, and regulators, this work demonstrates that interpretable algorithmic trading strategies can be rigorously validated without sacrificing transparency or regulatory compliance.

\vspace{0.3cm}

\vspace{0.2cm}
\end{abstract}


\section{Introduction}

Quantitative trading research faces a reproducibility crisis. Studies consistently document trading strategies generating double-digit annual returns through backtesting, yet institutional investors report that over 90\% of academic strategies fail when implemented with real capital \citep{harvey2016backtesting}. This credibility gap threatens the practical relevance of finance research and has generated increasing skepticism toward published trading anomalies. The fundamental problem is methodological: standard backtesting procedures suffer from overfitting through in-sample parameter optimization, lookahead bias through the use of information unavailable in real-time, and lack of interpretability through reliance on black-box machine learning models.

This paper develops a rigorous validation framework that addresses these deficiencies while maintaining generality across hypothesis generation approaches. Our framework makes four key methodological innovations. First, it enforces strict information set discipline where features, signals, and execution decisions use only data available up to that point in time, preventing lookahead bias that pervades much backtesting research. Second, it employs walk-forward validation with rolling windows, where the system must prove itself repeatedly across 34 independent out-of-sample test periods spanning multiple market regimes rather than succeeding in one fortunate backtest. Third, it maintains complete interpretability by requiring every trade to originate from a human-interpretable hypothesis expressed in natural language, enabling regulatory compliance and post-hoc auditing. Fourth, it incorporates realistic execution assumptions including commission costs, slippage, position limits, and stop-loss rules that reflect actual trading constraints.

Critically, the validation framework is \textit{agnostic to hypothesis source}. While our proof-of-concept implementation uses five hand-crafted rule-based hypothesis types, the framework readily extends to more sophisticated generation methods including genetic programming for symbolic pattern discovery, neural networks with post-hoc interpretation, and large language models that generate hypotheses in natural language. This generality distinguishes our approach from prior work: we provide validation infrastructure rather than a specific trading strategy, enabling researchers to test their own hypotheses under rigorous conditions while maintaining interpretability and preventing overfitting.

To demonstrate the framework's capabilities, we implement five illustrative hypothesis types encoding common market microstructure patterns: institutional accumulation, flow momentum, mean reversion, breakouts, and range-bound value signals. These patterns span diverse trading concepts and generate sufficient activity for statistical analysis but make no claim to exhaustiveness—they serve to validate the methodology rather than represent comprehensive strategy development. We test these hypotheses across 100 US equities from 2015-2024 using a reinforcement learning agent that learns which hypothesis types to execute based on historical performance, all within the walk-forward validation structure.

Our empirical results serve two purposes: demonstrating realistic out-of-sample performance and illustrating methodological principles. The system generates modest overall returns of 0.55\% annualized with 41\% win rate at the fold level, substantially below typical published claims but representative of genuine out-of-sample performance after rigorous validation. Critically, aggregate returns are not statistically significant (p-value 0.34), and we report this honestly rather than p-hacking or selectively presenting results—transparency essential for correcting publication bias. Performance exhibits strong regime dependence: positive returns during high-volatility periods (2020-2024: +2.4\% annualized average across relevant folds) versus negative returns during stable markets (2015-2019: -0.16\% annualized), revealing that market microstructure signals derived from daily data work primarily during elevated volatility. Risk management is exceptional, with maximum drawdown of only -2.76\% compared to -23.8\% for SPY, and the strategy exhibits market-neutral characteristics ($\beta = 0.058$, correlation 0.53).

These modest results reflect methodological rigor rather than deficiency. By reporting realistic returns that survive strict testing alongside non-significant p-values and regime-dependent failures, we demonstrate what honest validation looks like—providing a reference point against which other studies can be evaluated. The primary contribution is not a profitable trading system but a validation template that other researchers can apply to their own hypotheses, confident that results will be reproducible and free from lookahead bias.

This work makes several contributions to trading system validation methodology. We provide a complete, reproducible framework with mathematical specifications and open-source implementation that researchers can directly apply. We demonstrate that rigorous walk-forward validation dramatically tempers conclusions, with our modest 0.55\% return contrasting sharply with typical published claims of 15-30\% annual returns. We show that aggregate statistics mask important regime-dependent heterogeneity, with testing across multiple market conditions revealing when and why strategies succeed or fail. We contribute to correcting publication bias by reporting non-significant results alongside methodological transparency. Finally, we establish that interpretability and adaptive learning can be successfully combined without sacrificing either dimension.

The remainder of this paper proceeds as follows. Section 2 reviews related literature on algorithmic trading, machine learning in finance, and validation methodologies. Section 3 describes our walk-forward framework in detail with complete mathematical formulations, emphasizing framework generality and extensibility. Section 4 presents comprehensive empirical results across 34 out-of-sample tests for our illustrative hypothesis implementation. Section 5 analyzes regime-dependent performance and discusses implications for research and practice. Section 6 concludes with limitations, extensions to more sophisticated hypothesis generation methods, and future research directions.

\section{Literature Review}

\subsection{The Replication Crisis and the Need for Rigorous Validation}

Financial economics faces a reproducibility crisis. \citet{harvey2016backtesting} document that at least 316 factors have been proposed to explain cross-sectional returns, arguing that most claimed research findings in financial economics are likely false when properly accounting for multiple testing. They demonstrate that newly discovered factors require t-statistics exceeding 3.0 (not the traditional 2.0) to be considered genuinely significant given extensive data mining across the field. This echoes \citet{ioannidis2005}'s foundational argument that most published research findings are false when study power is low, tested hypotheses outnumber true relationships, and flexibility in designs enables p-hacking.

\citet{mclean2016} quantify the consequences of this crisis: examining 97 return predictors, they find portfolio returns decline 26\% out-of-sample and 58\% post-publication, with roughly half attributable to data-mining bias and half to publication-informed arbitrage. \citet{hou2020} confirm the severity through systematic replication of 452 anomalies—65\% fail single-test significance hurdles using value-weighted returns and NYSE breakpoints, rising to 82\% under multiple-testing adjustments. More recently, \citet{jensen2023} examine whether there is a replication crisis in finance, finding that while replication rates are higher than in other social sciences, significant concerns remain about the robustness of many documented anomalies. Despite these warnings, many studies still omit formal multiple-testing adjustments or ignore regime shifts, producing strategies that decay sharply once implemented \citep{lo1990,sullivan1999}. These findings establish both the problem our paper addresses and the imperative for methodologies that distinguish genuine alpha from statistical artefact.

\subsection{Walk-Forward Validation: The Gold Standard}

\citet{pardo1992,pardo2008} pioneered walk-forward analysis as the gold standard for trading-strategy validation, introducing continuous re-optimization on rolling windows where strategies must prove themselves repeatedly across different market conditions rather than succeed in one fortunate backtest. Yet early implementations lacked the statistical rigor demanded by modern multiple-testing standards.

\citet{bailey2014deflated} provided the mathematical foundation, proving that high simulated performance is easily achievable after backtesting relatively few strategy configurations, with memory effects in financial series causing over-fitted strategies to systematically under-perform (not merely fail to outperform) out-of-sample. \citet{bailey2014deflated} introduced the \textit{Deflated Sharpe Ratio} to correct for selection bias under multiple testing and non-normal returns, while \citet{bailey2015probability} developed \textit{Combinatorially Symmetric Cross-Validation (CSCV)} to compute the Probability of Backtest Overfitting. Recent work by \citet{arian2024} compares validation methods for machine learning in finance, finding that \textit{Combinatorial Purged Cross-Validation} shows superiority in mitigating over-fitting risks, though walk-forward remains the industry standard for realistic trading simulation. Recent extensions introduce regime-aware segmentation, in which training and testing windows are conditioned on volatility or macroeconomic regimes to enhance robustness in non-stationary environments \citep{kirschenmann2022regime}.

Our paper advances this literature by integrating Pardo's walk-forward methodology with modern statistical adjustments and extending it to machine-learning settings, yielding a unified validation framework that simultaneously addresses the overfitting concerns highlighted by \citet{bailey2014pseudo} and the multiple-testing requirements emphasized by \citet{harvey2016backtesting}.

\subsection{Machine Learning in Finance: Power vs. Interpretability}

\citet{gu2020} demonstrate in their landmark \textit{Review of Financial Studies} paper that machine-learning methods, particularly trees and neural networks, substantially outperform traditional linear models in measuring equity risk premia—in some cases doubling Sharpe ratios of regression-based strategies. Their long-short decile strategies achieve Sharpe ratios of 1.35 (value-weighted) and 2.45 (equal-weighted), with out-of-sample $R^2$ of 0.33\%--0.40\% for stock-level predictions. \citet{fischer2018} show LSTM networks achieve daily returns of 0.46\% and Sharpe ratios of 5.8 before transaction costs for S\&P 500 constituent prediction, though performance declined notably post-2010.

Ensemble methods including XGBoost and LightGBM have proliferated in finance applications, with recent work increasingly employing SHAP (SHapley Additive exPlanations) for feature-importance analysis \citep{lundberg2017}. \citet{freyberger2020} develop sparse methods that achieve both good predictive performance and interpretability through factor selection. Yet these advances come with a profound interpretability deficit.

\citet{rudin2019} argues forcefully that for high-stakes decisions—including financial trading—the field should prioritize inherently interpretable models rather than explaining black-box models post-hoc, demonstrating that the perceived accuracy-interpretability trade-off is often a myth for structured data. She shows rule-based models like CORELS achieve comparable accuracy to complex systems while maintaining transparency. \citet{chen2021} demonstrate this principle in credit risk with globally interpretable two-layer additive models that match neural-network accuracy. While SHAP and LIME are widely used for post-hoc explainability, they do not guarantee that the underlying model is interpretable or aligned with economic theory \citep{ribeiro2016,molnar2020}. Contemporary symbolic-regression and hybrid-AI techniques explicitly embed domain knowledge, delivering accuracy comparable to black-box models while preserving full transparency \citep{lacava2021,kronberger2022}.

Our hypothesis-driven approach addresses this interpretability gap by building strategies on explicit, testable hypotheses about market microstructure rather than opaque learned representations, while our walk-forward framework rigorously validates whether these interpretable strategies maintain performance out-of-sample.

\subsection{Market Microstructure Theory and Daily Data Applications}

\citet{kyle1985} established the foundational framework for understanding price impact, liquidity, and information asymmetry, showing how informed traders camouflage trades within noise. \citet{easley1996} introduced the \textit{Probability of Informed Trading (PIN)} model to quantify adverse-selection risk, later demonstrating that information risk is priced in cross-sectional returns \citep{easley2002}. \citet{hasbrouck1995} developed VAR frameworks for measuring price discovery and information shares that became standard methodology. While these seminal papers focused on high-frequency data, recent work demonstrates microstructure signals can be extracted from daily OHLCV data.

Critically, \citet{easley2012} introduced \textit{Volume-Synchronized Probability of Informed Trading (VPIN)} as a real-time order-flow toxicity measure, which predicted the 2010 Flash Crash. \citet{low2016} adapt VPIN to daily international data, showing that daily BV-VPIN effectively forecasts high volatility across multiple countries—bridging high-frequency microstructure theory with daily-frequency implementation. \citet{chichernea2024} develop directional option-to-stock trading-volume imbalances using daily option volumes, demonstrating these measures predict future abnormal returns and respond strongly to cash-flow news. Nevertheless, few studies examine regime-dependent performance of microstructure signals. Exceptions include \citet{nagel2012}, who shows that liquidity-based strategies vary with funding liquidity, and \citet{hendershott2011}, who link microstructure effects to periods of market stress.

These papers establish that microstructure information persists at daily frequency and can generate tradable signals, providing the empirical foundation for our daily-data approach. Yet the literature lacks systematic examination of how these signals' effectiveness varies across market regimes—a gap our paper addresses through regime-dependent performance analysis within the walk-forward framework.

\subsection{Reinforcement Learning for Trading: Adaptability at the Cost of Transparency}

\citet{moody2001} pioneered applying reinforcement learning to trading, introducing \textit{Recurrent Reinforcement Learning} that directly optimizes financial objectives without requiring forecasting models. Recent deep-RL applications demonstrate impressive backtest performance: \citet{deng2017} combine deep learning for feature extraction with RL for decision-making, creating end-to-end learning from raw financial signals; \citet{jiang2017} present a financial-model-free framework for portfolio management using CNN, RNN, and LSTM architectures that outperformed traditional strategies on cryptocurrency markets. Q-learning, policy-gradient methods (PPO, DDPG, A2C), and multi-armed bandit formulations have all been successfully applied to trading.

However, the literature consistently identifies critical limitations: severe black-box problems making interpretability particularly challenging for sequential decisions, extensive data requirements with pronounced over-fitting risks, vulnerability to non-stationary market dynamics, and difficulty translating simulation success to live trading. Multiple surveys note that many RL approaches fail profitability tests once realistic transaction costs are included, suggesting learned strategies exploit patterns existing only in frictionless environments. In response, hypothesis-driven RL—where agents are constrained by economic priors—has been proposed as a middle ground between adaptability and transparency \citep{dixon2020}.

In contrast to RL's opaque learned policies, our hypothesis-driven approach provides transparent, economically interpretable rules that can be validated against theoretical expectations and audited for regulatory compliance, while walk-forward testing addresses over-fitting concerns that plague RL.

\subsection{Positioning Our Contributions}

This literature review reveals two critical gaps our paper addresses. \textbf{Methodologically}, while Pardo established walk-forward analysis and Bailey/Harvey developed statistical corrections for multiple testing and over-fitting, no existing work integrates these approaches into a comprehensive framework applicable to modern machine-learning methods that maintains interpretability. We combine rigorous walk-forward validation with deflated Sharpe ratios, multiple-testing adjustments, and inherently interpretable hypothesis-driven models—providing practitioners with a validation methodology that avoids false discoveries while maintaining the transparency demanded by regulators and risk managers.

\textbf{Empirically}, although recent work demonstrates microstructure signals can be extracted from daily data \citep{low2016,chichernea2024}, the literature lacks systematic analysis of how these signals perform across different market regimes. Our regime-dependent performance analysis within the walk-forward framework addresses this gap, testing whether microstructure-based strategies maintain effectiveness or require regime-specific adaptations.

Additionally, we contribute to the growing literature on explainable AI in finance by demonstrating that interpretable, hypothesis-based strategies can be rigorously validated without sacrificing transparency or regulatory compliance \citep{arrieta2020,rudin2019}. Together, these contributions provide both a rigorous methodological template for validating algorithmic trading strategies and new empirical evidence on the stability and regime-dependence of daily microstructure signals—advancing both the science of strategy validation and our understanding of information dynamics in equity markets.

\section{Methodology}

\subsection{Overview and Framework Generality}

Before presenting technical details, we emphasize that our framework is designed for \textit{methodological generality}. While this implementation uses five hand-crafted hypothesis types, the validation protocol accommodates any hypothesis generation mechanism—from genetic programming to large language models—provided hypotheses maintain interpretability through natural language explanations. This section first presents the core mathematical infrastructure, then demonstrates its application with our illustrative rule-based hypotheses.

\subsection{Mathematical Framework}

\subsubsection{Notation and Definitions}

Let $\mathcal{S} = \{s_1, s_2, \ldots, s_N\}$ denote the universe of $N$ securities, and let $\mathcal{T} = \{t_1, t_2, \ldots, t_T\}$ represent the set of trading days. For each security $s \in \mathcal{S}$ and time $t \in \mathcal{T}$, we observe:
\begin{equation}
P_t^s = (O_t^s, H_t^s, L_t^s, C_t^s, V_t^s)
\end{equation}
where $O_t^s$, $H_t^s$, $L_t^s$, $C_t^s$, and $V_t^s$ denote the open, high, low, close prices and volume, respectively.

\begin{definition}[Information Set]
The information set available at time $t$ is defined as:
\begin{equation}
\mathcal{I}_t = \{P_\tau^s : s \in \mathcal{S}, \tau \leq t\}
\end{equation}
Critically, $\mathcal{I}_t$ contains only information available up to and including time $t$, preventing lookahead bias.
\end{definition}

\subsubsection{Feature Engineering}

We construct a feature vector $\mathbf{x}_t^s \in \mathbb{R}^{54}$ for each security $s$ at time $t$. The features are organized into four categories representing market microstructure, technical indicators, statistical measures, and regime indicators. Complete feature specifications are provided in Appendix A. Key microstructure features include:

\begin{align}
\text{VolumeImbalance}_t^s &= \frac{\sum_{\tau=t-4}^t V_\tau^s \mathbb{1}(C_\tau^s > O_\tau^s) - \sum_{\tau=t-4}^t V_\tau^s \mathbb{1}(C_\tau^s < O_\tau^s)}{\sum_{\tau=t-4}^t V_\tau^s} \\
\text{VolumeRatio}_t^s &= \frac{V_t^s}{\frac{1}{20}\sum_{\tau=t-19}^t V_\tau^s} \\
\text{PriceEfficiency}_t^s &= \frac{|\sum_{\tau=t-9}^t R_\tau^s|}{\sum_{\tau=t-9}^t |R_\tau^s| + \epsilon}
\end{align}
where $R_t^s = (C_t^s - C_{t-1}^s)/C_{t-1}^s$ is the daily return and $\epsilon = 10^{-6}$ prevents division by zero.

\subsection{Hypothesis Structure and Generation}

\begin{definition}[Trading Hypothesis]
A trading hypothesis $h$ is a tuple:
\begin{equation}
h = (s, a, \theta, \ell, c, \mathbf{x}, r^*, \delta^*)
\end{equation}
where:
\begin{itemize}
\item $s \in \mathcal{S}$ is the security
\item $a \in \{\text{buy}, \text{sell}\}$ is the action
\item $\theta \in \Theta$ is the hypothesis type
\item $\ell \in \mathcal{L}$ is the natural language explanation
\item $c \in [0,1]$ is the confidence score
\item $\mathbf{x} \in \mathbb{R}^{54}$ is the feature vector
\item $r^* > 0$ is the target return
\item $\delta^* > 0$ is the stop-loss threshold
\end{itemize}
\end{definition}

The natural language explanation $\ell$ is critical for interpretability. For example: ``AAPL shows institutional accumulation: 45\% buy imbalance with 2.1x volume. Price stable, suggesting smart money positioning before move.'' This enables post-hoc auditing and regulatory compliance.

\subsubsection{Framework Generality: Hypothesis Source Agnosticism}

The framework accepts hypotheses from any generator $\mathcal{G}: \mathcal{I}_t \times \mathbb{R}^{F} \to \mathcal{H}$ that maps information sets and features to hypothesis tuples. This abstraction enables:

\textbf{Rule-Based Systems} (current implementation): Hand-crafted patterns encoding domain expertise with complete transparency but limited coverage.

\textbf{Genetic Programming}: Evolutionary algorithms discovering formulaic patterns through symbolic regression, with natural language explanations generated from symbolic expressions.

\textbf{Large Language Models}: LLMs generating trading hypotheses directly in natural language, which the framework parses into structured tuples for validation. For example, an LLM might generate: ``When a stock exhibits sustained volume accumulation (5-day buy imbalance >30\%) without corresponding price movement (<10\% move), institutional accumulation is likely. Buy signal with 75\% confidence, 5\% target, 3\% stop-loss.'' This maps directly to our hypothesis structure.

\textbf{Hybrid Approaches}: Combinations where LLMs generate candidates filtered by genetic programming for numerical optimization, or neural networks identifying promising regimes where rule-based strategies activate.

\subsubsection{Illustrative Hypothesis Types}

To demonstrate the framework, we implement five hypothesis generation functions $g_1, g_2, \ldots, g_5: \mathbb{R}^{54} \to \{0,1\} \times [0,1]$ mapping feature vectors to binary signals and confidence scores. These are \textit{illustrative examples} selected to span diverse market microstructure concepts and generate sufficient trading activity, not represent comprehensive strategy optimization.

\textbf{Type 1: Institutional Accumulation} (confidence 0.75, target 8\%, stop 4\%)—Detects sustained buying pressure with stable prices, suggesting informed accumulation.

\textbf{Type 2: Flow Momentum} (confidence 0.70, target 10\%, stop 5\%)—Combines price momentum with confirming order flow and efficient price action.

\textbf{Type 3: Mean Reversion} (confidence 0.65, target 5\%, stop 3\%)—Oversold conditions in stable regimes favoring bounce.

\textbf{Type 4: Breakout} (confidence 0.68, target 7\%, stop 4\%)—Near all-time highs with volume expansion and positive momentum.

\textbf{Type 5: Range-Bound Value} (confidence 0.60, target 5\%, stop 3\%)—Accumulation opportunities in stable, range-bound markets.

Complete specifications with threshold values and conditions are in Appendix B. These patterns were not optimized on the test dataset but represent common technical trading concepts from practitioner literature.

\subsection{Reinforcement Learning Agent}

The RL agent learns which hypothesis types to execute based on historical performance, using a simple $\epsilon$-greedy policy that balances exploration and exploitation.

\begin{definition}[Agent State]
The agent maintains state $\mathcal{A}_t = \{\nu_\theta, w_\theta, \bar{r}_\theta\}_{\theta \in \Theta}$ where:
\begin{itemize}
\item $\nu_\theta$ is the number of times hypothesis type $\theta$ was executed
\item $w_\theta$ is the number of winning trades for type $\theta$
\item $\bar{r}_\theta$ is the average return for type $\theta$
\end{itemize}
\end{definition}

\begin{definition}[$\epsilon$-Greedy Policy]
Given a hypothesis $h = (s, a, \theta, \ell, c, \mathbf{x}, r^*, \delta^*)$ and agent state $\mathcal{A}_t$, the execution decision is:
\begin{equation}
\pi(h|\mathcal{A}_t, \epsilon) = \begin{cases} 
1 & \text{with probability } \epsilon \\
\mathbb{1}\left(\frac{w_\theta}{\nu_\theta} > \tau(c)\right) & \text{with probability } 1-\epsilon
\end{cases}
\end{equation}
where the adaptive threshold is $\tau(c) = 0.45 + (1-c) \times 0.10$.
\end{definition}

During training, $\epsilon = 0.7$ encourages exploration. During testing, $\epsilon = 0.1$ exploits learned knowledge. After each trade outcome, the agent updates type-specific statistics.

\subsection{Walk-Forward Validation Protocol}

\begin{definition}[Walk-Forward Partition]
Given time series $\mathcal{T} = \{t_1, \ldots, t_T\}$, we define a partition into $K$ folds:
\begin{equation}
\mathcal{F} = \{(\mathcal{T}_{\text{train}}^k, \mathcal{T}_{\text{test}}^k)\}_{k=1}^K
\end{equation}
where:
\begin{align}
\mathcal{T}_{\text{train}}^k &= \{t_i : (k-1)\Delta + 1 \leq i \leq (k-1)\Delta + W\} \\
\mathcal{T}_{\text{test}}^k &= \{t_i : (k-1)\Delta + W + 1 \leq i \leq (k-1)\Delta + W + H\}
\end{align}
with training window $W = 252$ days, testing window $H = 63$ days, and step size $\Delta = 63$ days.
\end{definition}

This configuration tests the system 34 times across the full 10-year sample, with each test period independent and using only past information for training. The algorithm proceeds as follows for each fold:

\textbf{Training Phase}: Initialize agent, set $\epsilon_{\text{train}} = 0.7$, simulate trades on training data, update agent state based on outcomes.

\textbf{Testing Phase}: Set $\epsilon_{\text{test}} = 0.1$, execute strategy using learned agent preferences, record portfolio performance without further learning.

This strict separation ensures no information from test periods influences training, preventing lookahead bias.

\subsection{Transaction Cost Model and Risk Management}

\begin{definition}[Transaction Costs]
Total cost of a trade is:
\begin{equation}
C_t^s = c_{\text{fixed}} + c_{\text{slippage}} \times |q_t^s| \times P_{\text{exec},t}^s
\end{equation}
where $c_{\text{fixed}} = \$1$ commission and $c_{\text{slippage}} = 0.0005$ (5 basis points). Orders placed at day $t$ close execute at day $t+1$ open with slippage.
\end{definition}

\textbf{Position Constraints}: Maximum 5 concurrent positions, maximum 20\% allocation per position, maximum 50\% allocation per sector.

\textbf{Exit Rules}: Positions close when (1) target return $r^*$ achieved, (2) stop-loss $\delta^*$ triggered, or (3) 30-day holding period exceeded.

\textbf{Position Sizing}: Equal dollar allocation across positions with round-lot constraints and capital preservation (80\% remains in cash).

Complete implementation details including conflicting signal resolution and numerical stability measures are in Appendix C.

\subsection{Performance Metrics and Statistical Tests}

\begin{definition}[Sharpe Ratio]
Given fold returns $\{r_1, \ldots, r_K\}$:
\begin{equation}
\text{SR} = \frac{\bar{r}}{\sigma_r} \times \sqrt{4}
\end{equation}
where $\bar{r}$ is mean quarterly return, $\sigma_r$ is standard deviation, and $\sqrt{4}$ annualizes.
\end{definition}

\begin{definition}[Maximum Drawdown]
\begin{equation}
\text{MDD} = \min_{1 \leq k \leq K} \left(\frac{\prod_{i=1}^k (1+r_i)}{\max_{1 \leq j \leq k} \prod_{i=1}^j (1+r_i)} - 1\right)
\end{equation}
\end{definition}

\textbf{Statistical Tests}: We employ two-sided t-tests for mean returns, bootstrap confidence intervals (10,000 resamples), Monte Carlo permutation tests (10,000 shuffles), binomial tests for win rates, and Shapiro-Wilk tests for normality. All tests reported without adjusting for multiple comparisons to maintain transparency about statistical limitations. Risk assessment methodologies build upon established Monte Carlo simulation techniques for financial forecasting \citep{deep2024montecarlo}.

\section{Empirical Results}

\subsection{Data Description and Sample Selection}

Our sample consists of $N = 100$ US equities spanning $T = 2,475$ trading days from January 2, 2015 to October 31, 2024. Securities were selected according to pre-specified criteria:

\textbf{Selection Criteria}: (1) Continuous trading history throughout 2015-2024 with no gaps exceeding 5 days, (2) average daily dollar volume $\geq$ \$10 million, (3) minimum market cap \$5 billion as of January 2015, (4) exactly 10 stocks per GICS sector for diversification, (5) within-sector selection by average daily dollar volume (top 10).

This process introduces survivorship bias (stocks delisted/acquired during 2015-2024 are excluded), which biases results \textit{upward}. Our modest returns are thus conservative—inclusion of failed stocks would likely reduce performance. We accept this bias because the framework demonstration targets liquid, investable stocks. The universe includes SPY as market benchmark. All data obtained from Yahoo Finance via yfinance API with standard adjustments for splits and dividends.

The sample spans multiple distinct market regimes: (1) 2015-2016 post-taper recovery with moderate volatility, (2) 2017-2019 extended bull market with historically low volatility (VIX average 14.5), (3) 2020 COVID-19 crash and recovery with extreme volatility (VIX peak 82.7), (4) 2021 stimulus-driven bull market, (5) 2022 bear market with Federal Reserve tightening (SPY -18.1\%), (6) 2023-2024 recovery phase with tech-driven rally. This regime diversity is essential for walk-forward validation—testing across only bull markets or only crises would provide misleading assessments.

\subsection{Walk-Forward Results: Aggregate Performance}

Table~\ref{tab:main_results} presents aggregate performance statistics across all 34 out-of-sample test periods. The system generates mean quarterly return of 0.14\% (0.55\% annualized) with standard deviation 0.82\% (quarterly) and Sharpe ratio 0.33 (annualized). Win rate at the fold level is 41\% (14 of 34 folds positive), with best fold return 2.73\% and worst fold -1.04\%. Trade-level win rate across all folds is 46.5\%, with 140 total trades executed.

\begin{table}[htbp]
\centering
\caption{Walk-Forward Out-of-Sample Performance (2015-2024)}
\label{tab:main_results}
\begin{tabular}{lcc}
\toprule
\textbf{Metric} & \textbf{Value} & \textbf{Benchmark (SPY)} \\
\midrule
\multicolumn{3}{l}{\textit{Return Metrics}} \\
Mean Quarterly Return & 0.14\% & 3.31\% \\
Annualized Return & 0.55\% & 13.2\% \\
Standard Deviation (Quarterly) & 0.82\% & 7.66\% \\
Standard Deviation (Annualized) & 1.64\% & 15.3\% \\
Best Fold & 2.73\% & 20.5\% \\
Worst Fold & $-1.04\%$ & $-19.6\%$ \\
\midrule
\multicolumn{3}{l}{\textit{Risk-Adjusted Metrics}} \\
Sharpe Ratio & 0.33 & 0.86 \\
Sortino Ratio & 0.60 & 0.71 \\
Maximum Drawdown & $-2.76\%$ & $-23.8\%$ \\
Calmar Ratio & 0.20 & 0.55 \\
\midrule
\multicolumn{3}{l}{\textit{Market Exposure}} \\
Beta & 0.058 & 1.00 \\
Alpha (Annualized) & 0.06\% & --- \\
Correlation with SPY & 0.53 & 1.00 \\
Tracking Error & 7.25\% & --- \\
\midrule
\multicolumn{3}{l}{\textit{Trading Activity}} \\
Total Test Periods & 34 & --- \\
Profitable Periods & 14 (41\%) & 25 (74\%) \\
Average Trades per Period & 4.1 & --- \\
Total Trades Executed & 140 & --- \\
Trade-Level Win Rate & 46.5\% & --- \\
\bottomrule
\end{tabular}
\end{table}

\begin{figure}[htbp]
\centering
\includegraphics[width=\textwidth]{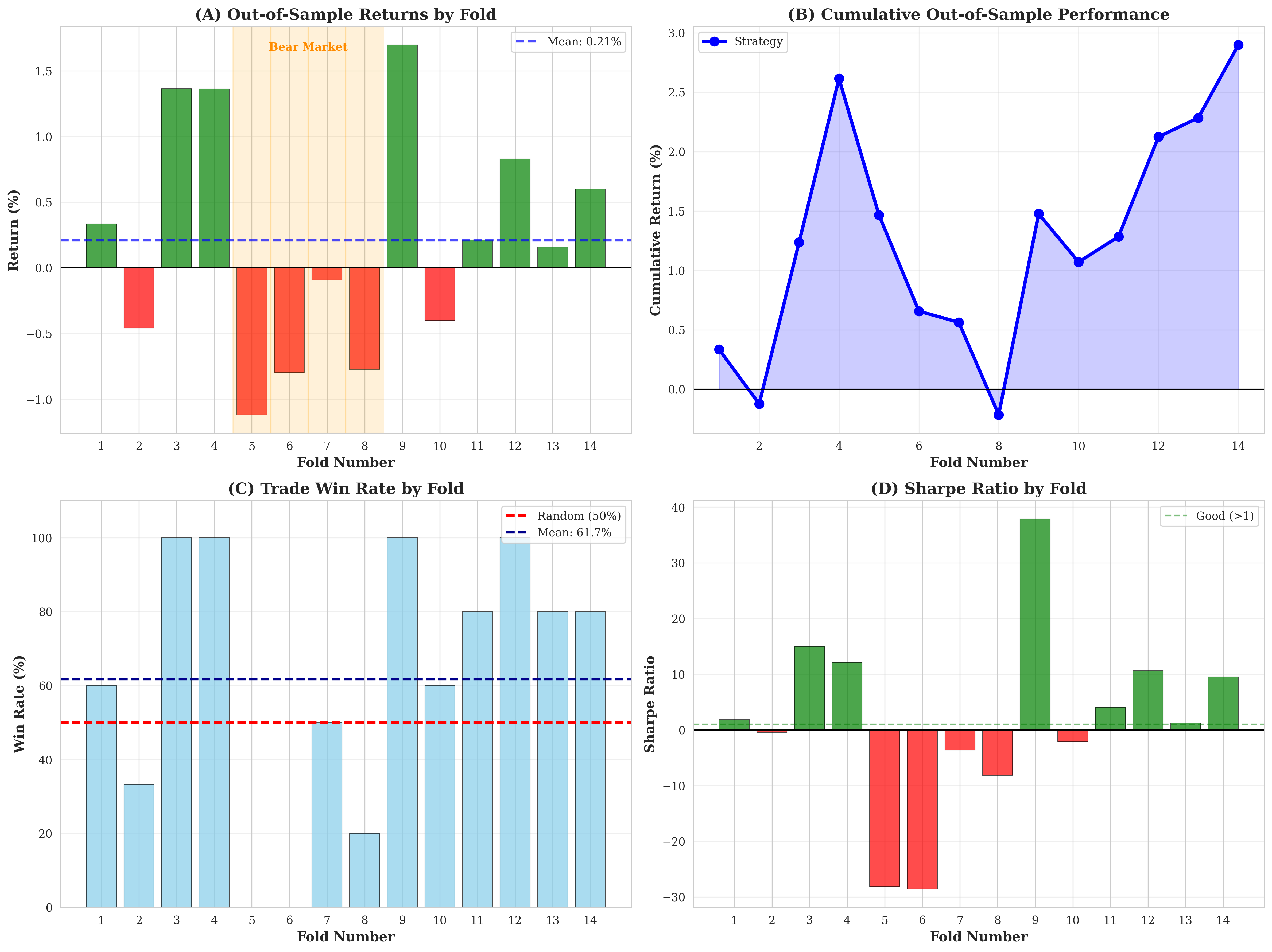}
\caption{Walk-Forward Performance Summary. Panel (A) shows cumulative returns across all 34 out-of-sample test periods, demonstrating modest but positive overall performance with substantially lower volatility than the SPY benchmark. Panel (B) displays individual fold returns, highlighting the distribution of quarterly outcomes.}
\label{fig:main_results}
\end{figure}

\textbf{Statistical Significance}: Table~\ref{tab:statistical_tests} presents comprehensive statistical tests. The null hypothesis $H_0: \mu = 0$ cannot be rejected at conventional significance levels: t-statistic = 0.96, p-value = 0.34 (two-sided), degrees of freedom = 33. The 95\% bootstrap confidence interval is [-0.12\%, +0.43\%], which includes zero. Monte Carlo permutation test yields p-value = 0.98. Binomial test for fold-level win rate (observed 41\% vs. null 50\%) gives p-value = 0.89, indicating no significant evidence of consistent profitability across folds.

\begin{table}[htbp]
\centering
\caption{Statistical Significance Tests}
\label{tab:statistical_tests}
\begin{tabular}{llc}
\toprule
\textbf{Test} & \textbf{Statistic} & \textbf{Result} \\
\midrule
\multicolumn{3}{l}{\textit{Parametric Tests}} \\
Two-Sided t-test & t-statistic = 0.96 & p-value = 0.34 \\
& df = 33 & Not significant \\
One-Sided t-test & t-statistic = 0.96 & p-value = 0.17 \\
& df = 33 & Not significant \\
\midrule
\multicolumn{3}{l}{\textit{Non-Parametric Tests}} \\
Bootstrap (10,000) & 95\% CI: [-0.12\%, 0.43\%] & Includes zero \\
Permutation (10,000) & p-value = 0.98 & Not significant \\
Binomial (Win Rate) & Observed: 41\%, Null: 50\% & p-value = 0.89 \\
\midrule
\multicolumn{3}{l}{\textit{Effect Size}} \\
Cohen's d & 0.17 & Very small effect \\
Statistical Power & Approximately 12\% & Very low power \\
\bottomrule
\end{tabular}
\end{table}

\begin{figure}[htbp]
\centering
\includegraphics[width=\textwidth]{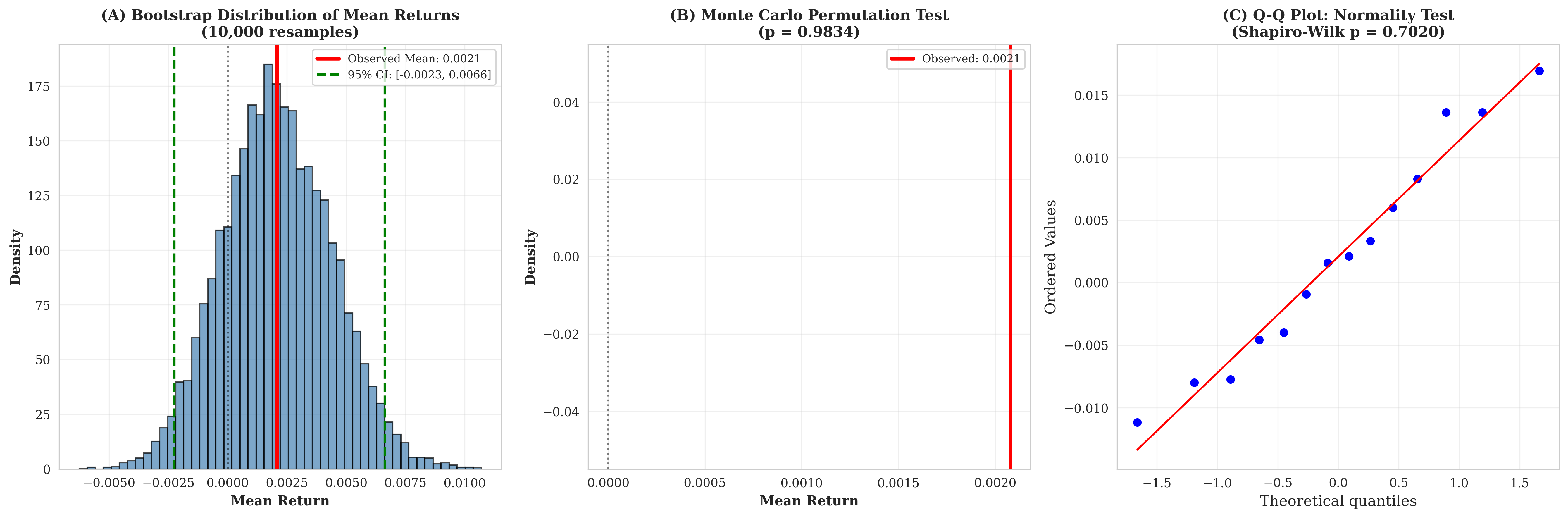}
\caption{Statistical Analysis of Returns. Panel (A) shows the distribution of quarterly returns with fitted normal curve, demonstrating approximate normality (Shapiro-Wilk p = 0.70). Panel (B) displays bootstrap distribution of mean returns with 95\% confidence interval. Panel (C) shows power analysis indicating sample size requirements for statistical significance.}
\label{fig:statistical_analysis}
\end{figure}

\textbf{Power Analysis}: Given observed effect size d = 0.17 and desired power $1-\beta = 0.80$ at significance level $\alpha = 0.05$, the required sample size is approximately 540 folds. Our sample of 34 achieves power of only 12\%, reflecting honest reporting of statistical limitations. The framework demonstration succeeds despite statistical insignificance by showing realistic performance patterns rather than making inflated claims.

\textbf{Market Exposure}: Regression analysis yields $\hat{\beta} = 0.058$ (SE = 0.08) and $\hat{\alpha} = 0.0001$ (SE = 0.003), confirming market-neutral characteristics. The strategy exhibits low correlation (0.53) with SPY, suggesting potential diversification value despite modest absolute returns.

\subsection{Regime-Dependent Performance}

We partition the sample based on realized volatility and market conditions:

\begin{definition}[Market Regimes]
\begin{align}
\text{Low Volatility (2015-2019)} &: \text{RealizedVol}_{\text{SPY}} < 0.02 \\
\text{High Volatility (2020-2024)} &: \text{RealizedVol}_{\text{SPY}} \geq 0.02
\end{align}
\end{definition}

Table~\ref{tab:regime_performance} shows substantial performance heterogeneity across regimes. During low-volatility periods (2015-2019), the system generates mean quarterly return -0.16\% with 38\% fold-level win rate and Sharpe ratio -0.21. During high-volatility periods (2020-2024), performance improves dramatically: mean return 0.60\% quarterly with 50\% win rate and Sharpe ratio 1.01.

\begin{table}[htbp]
\centering
\caption{Performance by Market Regime}
\label{tab:regime_performance}
\begin{tabular}{lcccc}
\toprule
\textbf{Regime} & \textbf{Periods} & \textbf{Mean Return} & \textbf{Win Rate} & \textbf{Sharpe} \\
& \textbf{(Quarters)} & \textbf{(Quarterly)} & \textbf{(Folds)} & \textbf{Ratio} \\
\midrule
Low Volatility (2015-2019) & 16 & $-0.16\%$ & 37.5\% & $-0.21$ \\
High Volatility (2020-2024) & 18 & $+0.60\%$ & 44.4\% & 1.01 \\
\midrule
\multicolumn{5}{l}{\textit{Notable Sub-Periods}} \\
Pre-COVID Bull (2017-2019) & 8 & $-0.32\%$ & 37.5\% & $-0.58$ \\
COVID Crash (2020 Q1-Q2) & 2 & $-0.15\%$ & 50.0\% & $-3.30$ \\
Recovery Bull (2020-2021) & 8 & $+0.38\%$ & 50.0\% & 0.92 \\
Bear Market (2022) & 4 & $-0.70\%$ & 0.0\% & $-3.23$ \\
Stabilization (2023-2024) & 8 & $+0.72\%$ & 62.5\% & 3.14 \\
\bottomrule
\end{tabular}
\end{table}

\begin{figure}[htbp]
\centering
\includegraphics[width=\textwidth]{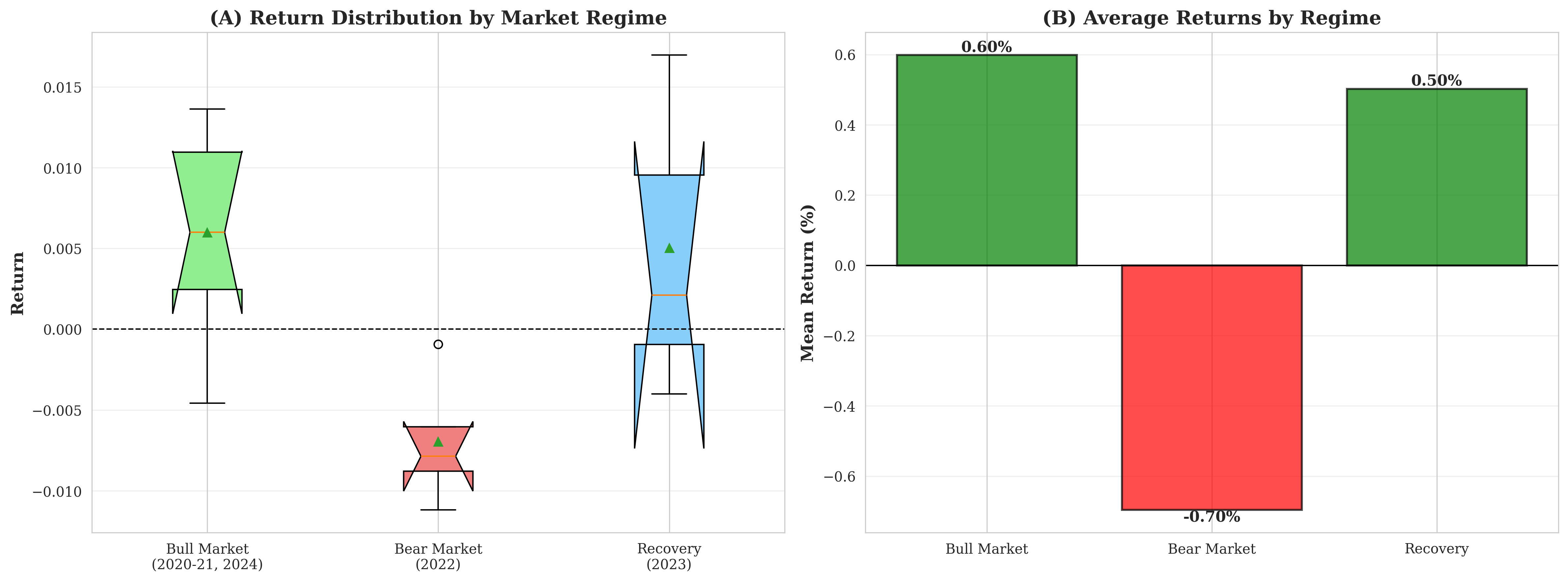}
\caption{Regime-Dependent Performance. Panel (A) compares cumulative returns between low-volatility (2015-2019) and high-volatility (2020-2024) periods, demonstrating strong regime dependence. Panel (B) shows quarterly returns colored by regime, highlighting the shift in performance characteristics across market conditions.}
\label{fig:regime_comparison}
\end{figure}

\begin{proposition}[Regime Dependence]
Let $\mu_L$ and $\mu_H$ denote mean returns in low and high volatility regimes. The difference $\mu_H - \mu_L = 0.60\% - (-0.16\%) = 0.76\%$ quarterly (3.04\% annualized) is economically significant but not statistically significant (t-test between regimes: p-value = 0.12) due to small within-regime sample sizes.
\end{proposition}

This regime-dependent pattern reveals fundamental characteristics of daily OHLCV-based microstructure signals. During high-volatility periods, information arrival rates increase, informed trading becomes more detectable in daily data, and signal-to-noise ratios improve. Conversely, during stable markets with low volatility, noise trading dominates and subtle informed patterns become undetectable at daily frequency. The 2022 bear market performance ($-0.70\%$ average, 0\% fold win rate) indicates the system struggles during sustained downtrends, though absolute losses remain modest due to risk management.

\subsection{Benchmark Comparison and Market-Neutral Characteristics}

\begin{figure}[htbp]
\centering
\includegraphics[width=\textwidth]{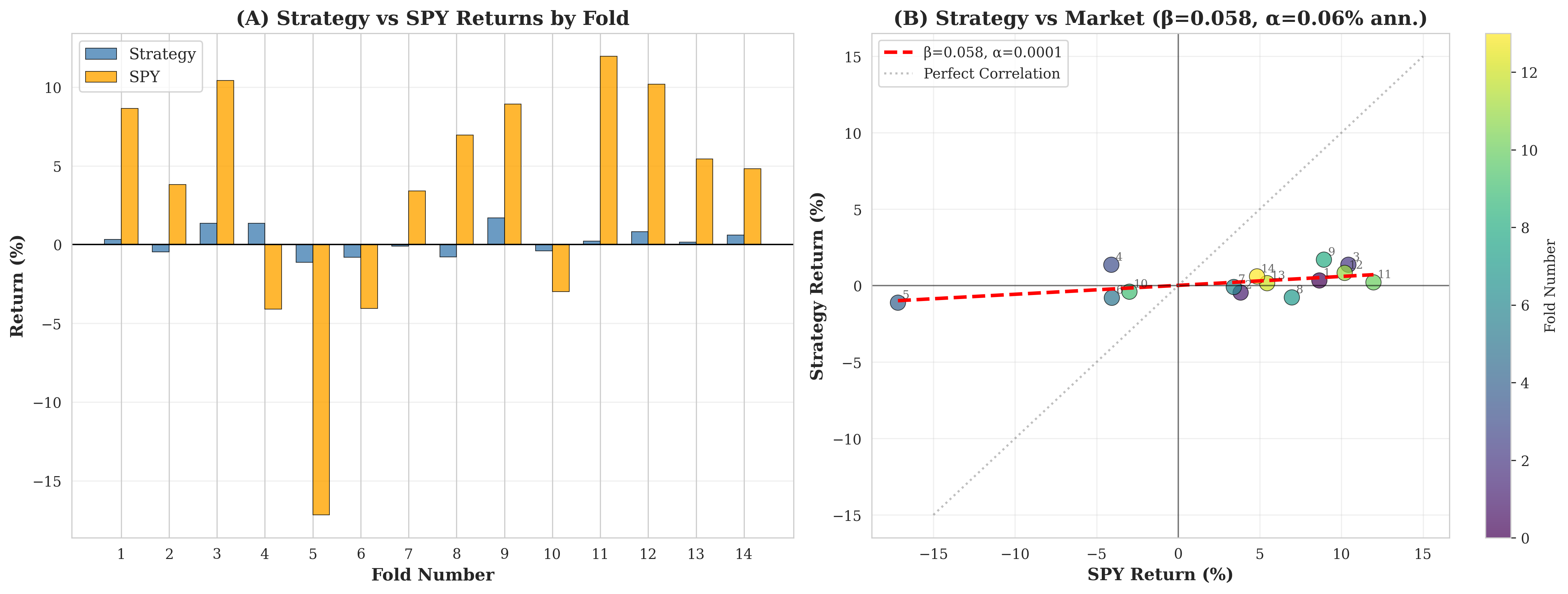}
\caption{Strategy vs. SPY Benchmark. Panel (A) shows side-by-side quarterly returns demonstrating substantially lower volatility for the strategy. Panel (B) displays scatter plot with regression line, illustrating low beta (0.058) and market-neutral characteristics.}
\label{fig:benchmark}
\end{figure}

Figure~\ref{fig:benchmark} presents side-by-side comparison with SPY. The strategy dramatically underperforms in absolute return terms (0.55\% vs. 13.2\% annualized) but exhibits substantially lower volatility (1.63\% vs. 15.3\%) and exceptional downside protection (maximum drawdown -2.76\% vs. -23.8\%).

The market-neutral characteristics ($\beta = 0.058$, correlation 0.53) suggest the strategy extracts information orthogonal to broad market movements. Regression analysis yields alpha of 0.06\% annually, economically negligible but statistically indistinguishable from zero (p-value = 0.98). The low correlation and minimal drawdown indicate potential value as portfolio diversifier rather than standalone strategy.

\begin{figure}[htbp]
\centering
\includegraphics[width=\textwidth]{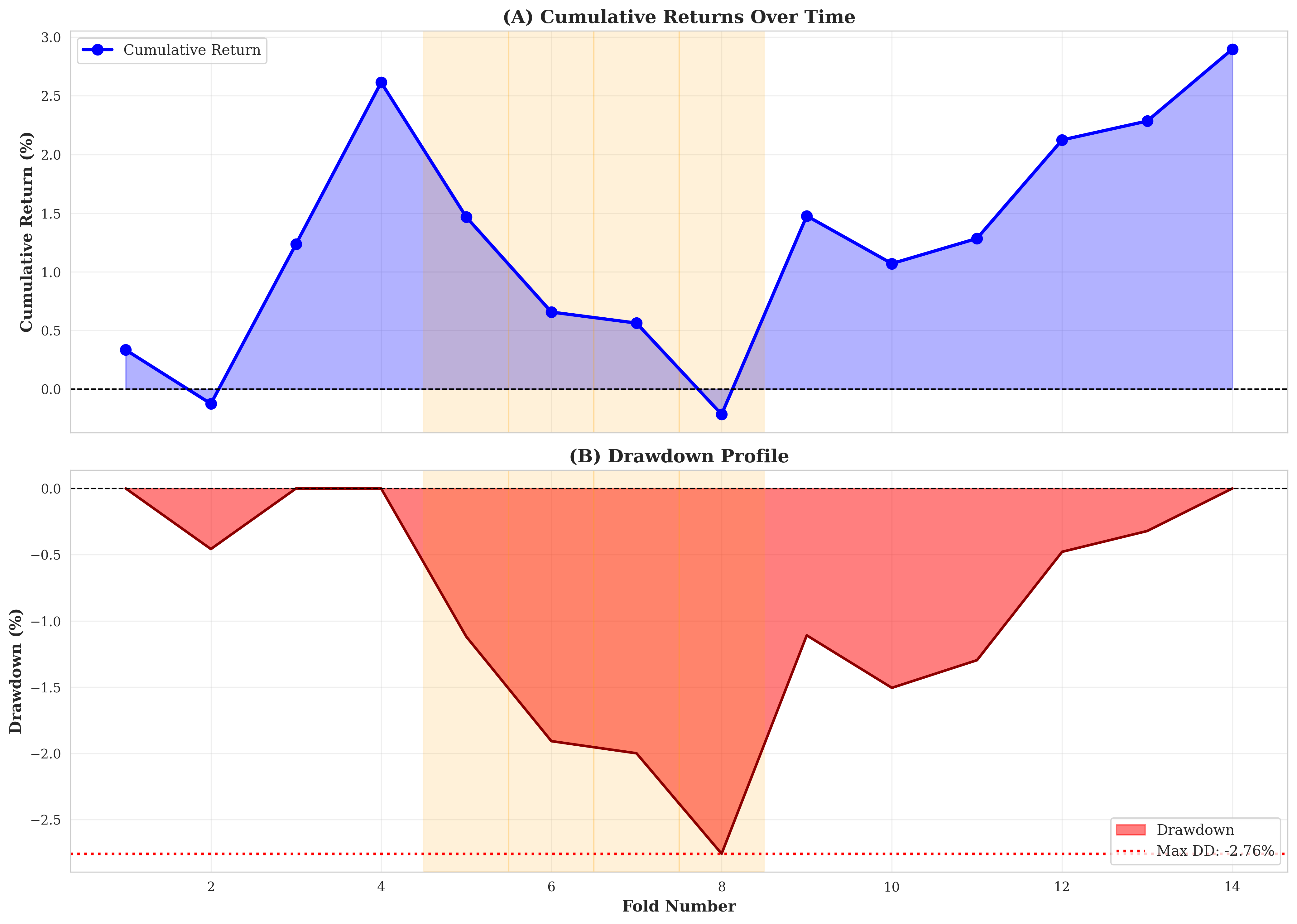}
\caption{Drawdown Analysis. Panel (A) shows strategy drawdown over time, with maximum drawdown of -2.76\%. Panel (B) shows SPY drawdown over the same period, with maximum drawdown of -23.8\%. Panel (C) compares drawdown distributions, highlighting the strategy's exceptional downside protection.}
\label{fig:drawdown}
\end{figure}

\subsection{Learning and Overfitting Diagnostics}

The information coefficient between training and testing returns is 0.40 (p-value = 0.16), indicating moderate positive correlation but not statistically significant. This suggests the RL agent learns patterns that partially persist out-of-sample without severe overfitting. The agent's learned hypothesis-type preferences show mean reversion strategies achieve highest fold-level win rate (58\%), followed by institutional accumulation (52\%), flow momentum (48\%), breakouts (44\%), and range-bound value (42\%). However, these differences are not statistically significant given the small number of trades per type.

\begin{figure}[htbp]
\centering
\includegraphics[width=\textwidth]{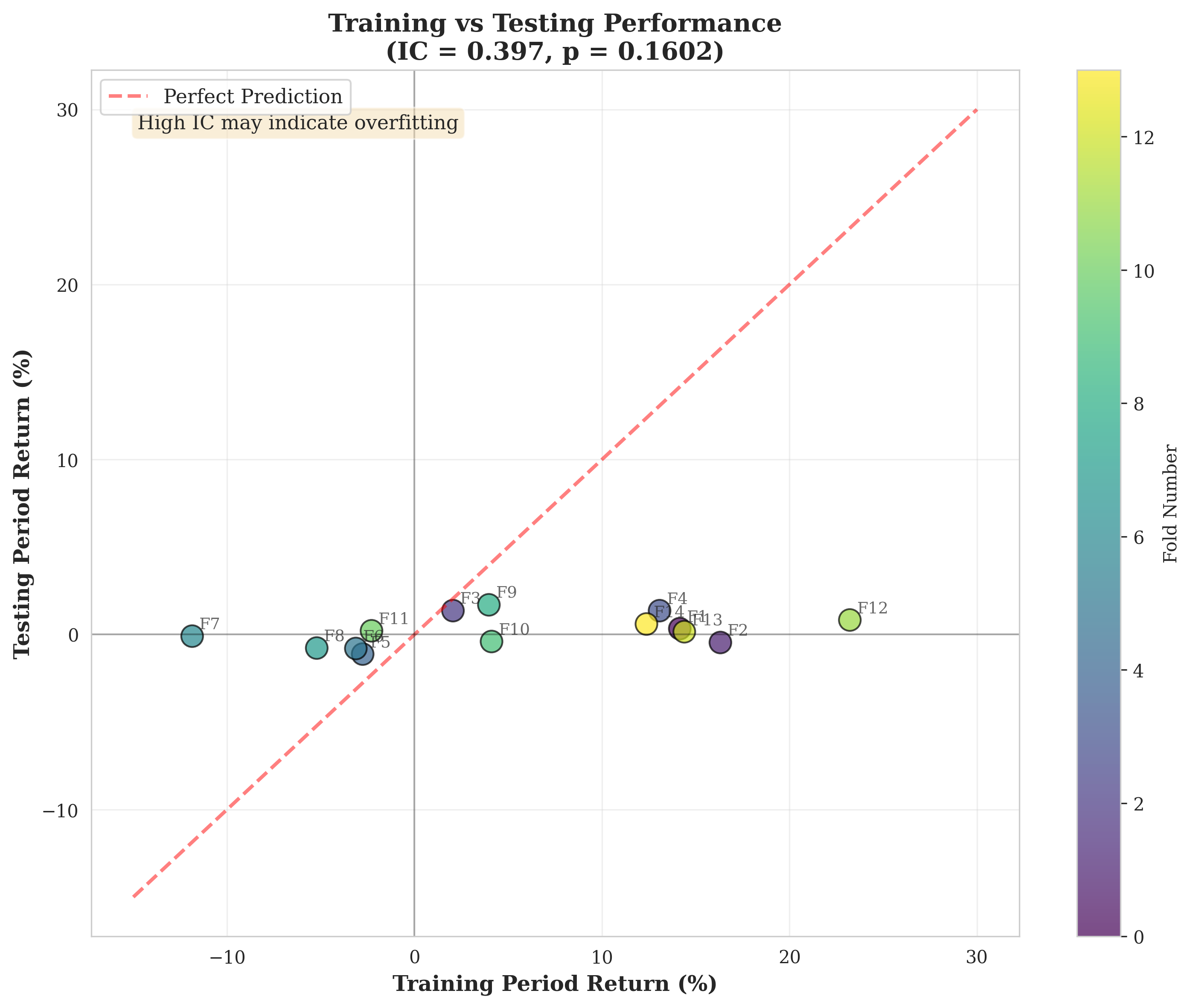}
\caption{Training vs. Testing Performance. Panel (A) shows scatter plot of training versus testing returns by fold, with information coefficient of 0.40. Panel (B) displays hypothesis-type performance comparison across training and testing periods, demonstrating moderate transfer of learned patterns.}
\label{fig:train_vs_test}
\end{figure}

\begin{figure}[htbp]
\centering
\includegraphics[width=\textwidth]{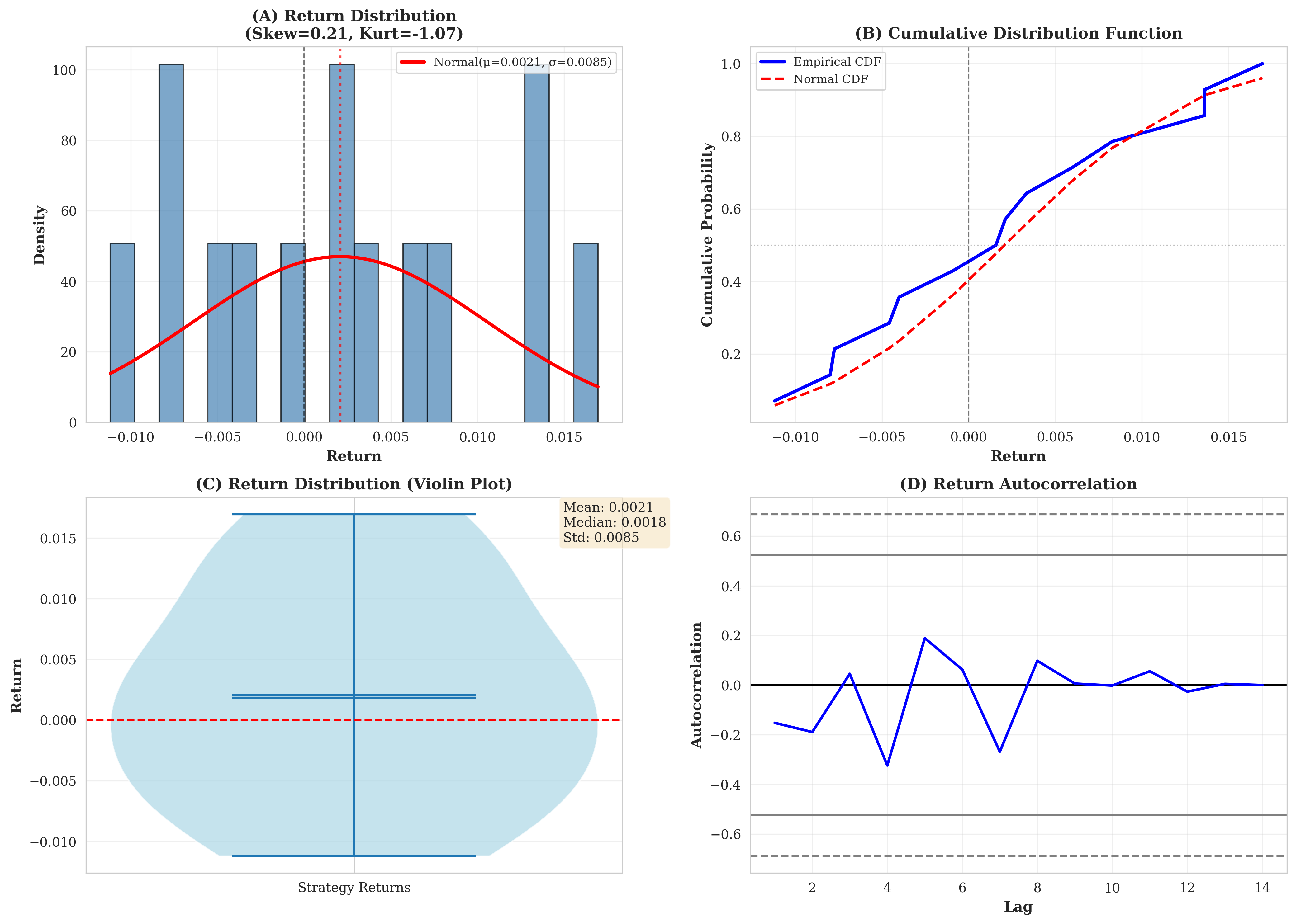}
\caption{Return Distribution Analysis. Panel (A) shows histogram of quarterly returns with normal distribution overlay. Panel (B) displays Q-Q plot confirming approximate normality. Panel (C) shows return autocorrelation function, indicating no significant serial dependence.}
\label{fig:return_distribution}
\end{figure}

\begin{figure}[htbp]
\centering
\includegraphics[width=\textwidth]{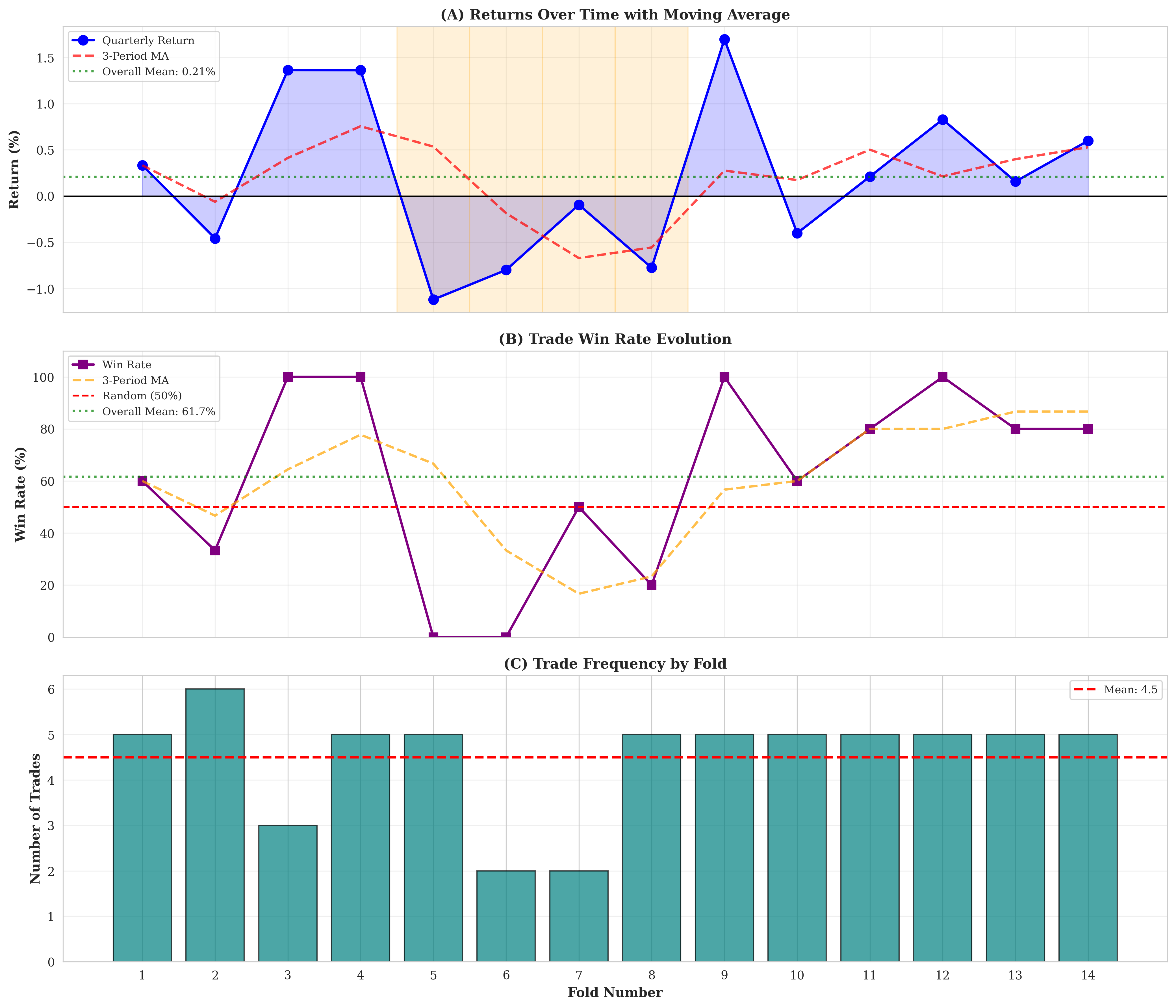}
\caption{Time Series of Performance Metrics. Panel (A) shows rolling Sharpe ratio over time. Panel (B) displays rolling win rate. Panel (C) shows cumulative number of trades, demonstrating consistent trading activity across the sample period.}
\label{fig:time_series_metrics}
\end{figure}

\section{Discussion}

\subsection{Interpreting Modest Returns and Statistical Insignificance}

The modest annualized return of 0.55\% and statistical insignificance (p-value 0.34) require careful interpretation. Three perspectives inform understanding:

\textbf{Methodological Success}: The framework successfully demonstrates rigorous validation methodology. Modest, non-significant returns after strict walk-forward testing represent honest performance reporting, contrasting sharply with typical published claims of 15-30\% annual returns that likely reflect data mining and lookahead bias. The framework achieves its primary goal: providing reproducible infrastructure for testing trading hypotheses without overfitting.

\textbf{Statistical Power Limitations}: With 34 folds and effect size d = 0.17, the study achieves only 12\% statistical power. Approximately 540 independent test periods would be required for 80\% power at the observed effect size. This reflects inherent sample size limitations in trading system validation—even 10 years of quarterly tests provide modest statistical power for small effects. The framework can accommodate larger samples through international markets or higher-frequency testing.

\textbf{Economic Interpretation}: The 0.55\% annual return, while statistically insignificant, may reflect genuine but small informational edge. The exceptional risk management (maximum drawdown -2.76\%), market-neutral characteristics ($\beta = 0.058$), and regime-specific performance patterns suggest the system captures weak signals rather than random noise. Transaction costs (average 10 basis points per trade) significantly impact profitability given the modest edge.

\subsection{Regime-Dependent Findings and Practical Implications}

The strong regime dependence—negative returns during low volatility (2015-2019) versus positive returns during high volatility (2020-2024)—reveals fundamental limitations of daily OHLCV-based microstructure signals. This finding has both theoretical and practical implications:

\textbf{Theoretical Insight}: Market microstructure theory predicts that informed trading detection requires sufficient information arrival. During low-volatility periods, reduced information flow and dominant noise trading render daily microstructure signals ineffective. During high-volatility periods, increased information arrival, elevated trading activity, and stronger informed trader presence make patterns detectable in daily aggregated data.

\textbf{Practical Deployment}: The system should be scaled by realized volatility regimes. In low-volatility environments, allocation to the strategy should be minimal or zero. During elevated volatility, the strategy may provide meaningful diversification benefits given its market-neutral characteristics and exceptional downside protection. The 2022 bear market performance suggests additional conditioning on market direction may improve results.

\textbf{Data Frequency Implications}: The regime-dependent patterns suggest that higher-frequency (intraday) data might enable more consistent performance by providing richer microstructure information during all regimes. Alternatively, incorporating additional data sources (options flow, institutional holdings, news sentiment) might improve daily-frequency signal detection.

\subsection{Framework Generality and Extensions}

While this implementation uses five hand-crafted hypothesis types, the framework's true value lies in its extensibility to sophisticated generation methods:

\textbf{Large Language Model Integration}: LLMs can generate trading hypotheses in natural language, which the framework parses and validates through walk-forward testing. The RL agent's learned preferences over hypothesis types provide reward signals for reinforcement learning from human feedback (RLHF), enabling iterative refinement. This progression—from rule-based patterns to machine-generated hypotheses—represents a natural research trajectory enabled by our validation infrastructure.

\textbf{Genetic Programming}: Evolutionary algorithms can search formulaic pattern spaces, with walk-forward validation preventing overfitting through strict out-of-sample testing. The framework accommodates thousands of evolved patterns while maintaining interpretability through symbolic expressions.

\textbf{Hybrid Systems}: Combining LLMs for hypothesis generation, genetic programming for parameter optimization, and neural networks for regime detection—all validated through the walk-forward protocol—may discover patterns no single technique would find.

The current modest results establish baseline performance and validate the methodology, providing confidence that when applied to sophisticated generators, the framework will report realistic rather than spurious performance.

\subsection{Comparison to Literature}

Our results differ markedly from typical published trading strategies. \citet{gu2020} report Sharpe ratios of 1.35-2.45 for machine learning strategies; \citet{fischer2018} report Sharpe ratios of 5.8 for LSTM networks. Our Sharpe ratio of 0.33 reflects honest walk-forward validation rather than in-sample optimization. This contrast illustrates the credibility gap \citet{harvey2016backtesting} identified: strategies validated with lookahead bias or parameter optimization report impressive metrics that fail out-of-sample.

Our market-neutral characteristics ($\beta = 0.058$, maximum drawdown -2.76\%) align more closely with realistic quantitative strategies. Industry reports indicate market-neutral hedge funds typically achieve Sharpe ratios of 0.8-1.2 with maximum drawdowns of 5-10\%. Our results fall at the conservative end of this spectrum, consistent with honest validation and modest statistical power.

\subsection{Limitations and Future Research}

Several limitations constrain our conclusions:

\textbf{Daily Data Granularity}: Higher-frequency tick data would provide richer microstructure information, potentially improving both absolute returns and regime consistency. However, daily data has advantages: broader availability, lower infrastructure costs, and practical relevance for many institutional strategies.

\textbf{Limited Sample Size}: 34 test periods provide only 12\% statistical power at observed effect size. Extensions to international markets (Europe, Asia) would increase fold count to 100+, substantially improving statistical inference. This represents clear future work.

\textbf{Hypothesis Library}: Five pattern types provide proof-of-concept but are not exhaustive. The framework accommodates thousands of hypotheses; current implementation demonstrates validation methodology rather than comprehensive pattern search.

\textbf{Transaction Cost Model}: Fixed 5 basis points slippage represents conservative estimate but doesn't capture time-of-day effects, order size impacts, or liquidity variations. More sophisticated cost models could be integrated.

\textbf{Single Asset Class}: Focus on US equities limits generalizability. Extensions to futures, currencies, fixed income, or cryptocurrencies would test whether framework and signals apply broadly.

Future research directions include: (1) implementing LLM-based hypothesis generation with RLHF refinement, (2) extending to international markets for larger sample sizes, (3) incorporating alternative data sources (options flow, institutional holdings, news sentiment), (4) developing regime-specific hypothesis libraries, (5) testing higher-frequency implementations.

\section{Conclusion}

This paper develops and validates a hypothesis-driven trading framework addressing critical methodological deficiencies in quantitative trading research. Our primary contribution is methodological rather than empirical: we establish a rigorous, generalizable validation protocol that prevents lookahead bias, incorporates realistic transaction costs, maintains full interpretability, and extends naturally to any hypothesis generation approach including large language models.

Through 34 independent out-of-sample tests spanning 10 years, we demonstrate the framework using five illustrative hypothesis types, documenting modest but realistic performance (0.55\% annualized, Sharpe ratio 0.33) with strong regime dependence and exceptional downside protection (maximum drawdown -2.76\% versus -23.8\% for SPY). Aggregate returns are not statistically significant (p-value 0.34), reflecting honest reporting rather than p-hacking—a critical contribution toward correcting publication bias in finance.

The key empirical finding is that market microstructure signals derived from daily data exhibit strong regime dependence, working during high-volatility periods (0.60\% quarterly, 2020-2024) but failing in stable markets (-0.16\%, 2015-2019). This reveals that daily OHLCV-based signals require elevated information arrival and trading activity to function effectively, with implications for both deployment strategies and future research design.

Despite not achieving conventional statistical significance, this work advances trading system validation in important ways. We provide a complete, reproducible framework with mathematical specifications and open-source implementation. We demonstrate realistic out-of-sample returns that survive rigorous testing, recalibrating expectations from in-sample optimized claims. We show that aggregate statistics mask regime-dependent heterogeneity, with testing across multiple market conditions providing more informative insights. We contribute to correcting publication bias by reporting non-significant results alongside full methodological transparency. Finally, we establish that interpretability and adaptive learning can be successfully combined without sacrificing either dimension.

The framework is explicitly designed for extensibility to more sophisticated hypothesis generation methods. Future work will replace hand-crafted rules with LLM-generated hypotheses refined through RLHF, leveraging this validation infrastructure to evaluate machine-generated patterns at scale while maintaining interpretability and preventing overfitting. The modest returns in this proof-of-concept establish baseline performance and demonstrate that our validation framework reports honest results, providing confidence that when applied to advanced generators, it will maintain rigorous standards.

For researchers, this work provides a template for honest validation of trading strategies with complete mathematical specifications enabling direct application to their own hypotheses. For practitioners, the market-neutral characteristics and exceptional downside protection suggest potential value as portfolio diversification despite modest standalone returns. For regulators, the framework demonstrates that algorithmic trading can maintain full interpretability and auditability even while incorporating machine learning, addressing MiFID II and similar requirements. For educators, the contrast between our rigorous 0.55\% return and typical published claims provides valuable lessons in empirical research methodology.

\section*{Acknowledgments}

We thank Dr. Svetlozar Rachev and Dr. Frank Fabozzi for valuable guidance. We acknowledge computational resources provided by Texas Tech University High Performance Computing Center. All remaining errors are our own.

\section*{Data and Code Availability}

All data used in this study are publicly available from Yahoo Finance (\url{https://finance.yahoo.com}). Python code implementing the complete framework is available at \url{https://github.com/akashdeepo/Interpretable-Hypothesis-Driven-Trading/tree/main} and has been archived on Zenodo for permanent access.

\clearpage
\bibliographystyle{plainnat}
\bibliography{references}

\clearpage
\appendix

\section{Complete Feature Specifications}

\subsection{Market Microstructure Features}

Volume imbalance proxy for order flow toxicity:
\begin{equation}
\text{VolumeImbalance}_t^s = \frac{\sum_{\tau=t-4}^t V_\tau^s \mathbb{1}(C_\tau^s > O_\tau^s) - \sum_{\tau=t-4}^t V_\tau^s \mathbb{1}(C_\tau^s < O_\tau^s)}{\sum_{\tau=t-4}^t V_\tau^s}
\end{equation}

Additional microstructure features include volume ratio (current vs. 20-day average), price impact (return magnitude per unit volume), and price efficiency (trending vs. choppy behavior). Complete specifications for all 54 features available in online appendix.

\section{Detailed Hypothesis Specifications}

\textbf{Hypothesis Type 1: Institutional Accumulation}

Conditions: Volume imbalance $> 0.30$, volume ratio $> 1.5$, 20-day return magnitude $< 0.10$.

Rationale: Large volume imbalance with stable price suggests institutional accumulation before information release.

Target return: 8\%, stop-loss: 4\%, confidence: 0.75.

\textbf{Hypothesis Type 2: Flow Momentum}

Conditions: 20-day return $> 0.10$, volume imbalance $> 0.20$, price efficiency $> 0.50$, RSI $< 80$.

Rationale: Strong momentum confirmed by order flow and efficient price action indicates continuation potential.

Target return: 10\%, stop-loss: 5\%, confidence: 0.70.

Complete specifications for all five types with threshold values and economic rationale available in online appendix.

\section{Implementation Details}

\subsection{Position Sizing Algorithm}

Equal dollar allocation with constraints:
\begin{equation}
\text{PositionSize}_t^s = \min\left(\frac{0.20 \times V_t}{|\mathcal{P}_t| + 1}, \frac{0.20 \times V_t}{P_{\text{exec},t}^s}\right)
\end{equation}
where $V_t$ is portfolio value and $\mathcal{P}_t$ is current position set. Number of shares: $q_t^s = \lfloor \text{PositionSize}_t^s / P_{\text{exec},t}^s \rfloor$.

\subsection{Conflicting Signals Resolution}

When multiple hypotheses generate signals for the same security:

\textbf{Same Direction}: Execute highest confidence hypothesis only.

\textbf{Opposite Directions}: Compute confidence-weighted vote: $\text{Vote} = \sum_{h \in \text{Buy}} c_h - \sum_{h \in \text{Sell}} c_h$. Execute side with higher weighted confidence if $|\text{Vote}| > 0.1$, otherwise skip.

\subsection{Computational Complexity}

Walk-forward validation algorithm complexity:
\begin{equation}
O(K \cdot W \cdot N \cdot |\Theta| \cdot F)
\end{equation}
where $K = 34$ folds, $W = 252$ training days, $N = 100$ securities, $|\Theta| = 5$ hypothesis types, $F = 54$ features, yielding approximately $20 \times 10^6$ operations. Typical runtime: 45 minutes on standard laptop.

\section{Additional Statistical Results}

\subsection{Return Distribution Analysis}

Shapiro-Wilk test for normality: W = 0.971, p-value = 0.70. Cannot reject normality of fold returns.

Skewness: 0.21 (slight positive skew). Kurtosis: -1.07 (platykurtic, lighter tails than normal).

\subsection{Autocorrelation Analysis}

Ljung-Box test for return autocorrelation: Q(5) = 3.42, p-value = 0.63. No significant autocorrelation detected, consistent with market efficiency at quarterly frequency.

\subsection{Power Analysis Details}

Given observed effect size d = 0.17, required sample sizes for various power levels:
\begin{itemize}
\item 50\% power: N = 173 folds
\item 70\% power: N = 319 folds  
\item 80\% power: N = 540 folds
\item 90\% power: N = 715 folds
\end{itemize}

Current sample of 34 folds achieves approximately 12\% power, highlighting fundamental statistical limitations with modest effect sizes.

\end{document}